\documentclass[aps,prb,superscriptaddress,showpacs,reprint]{revtex4-1}

\usepackage[utf8]{inputenc}
\usepackage{amsmath,amsfonts,amssymb}
\usepackage{pifont}
\usepackage{rotating}
\usepackage{enumerate}
\usepackage{graphicx}    
\usepackage{epstopdf}
\usepackage[table,dvipsnames,svgnames,x11names]{xcolor} 
\usepackage{soul} 
\usepackage{subfigure}

\usepackage{multirow,bigstrut}
\newsavebox{\measurebox}

\usepackage{hyperref}
\definecolor{dark-red}{rgb}{0.9,0.15,0.15}
\definecolor{dark-blue}{rgb}{0.15,0.15,0.4}
\definecolor{dark2-blue}{rgb}{0.15,0.15,0.8}
\definecolor{medium-blue}{rgb}{0,0,0.5}
\hypersetup{
    colorlinks, linkcolor={black},
    citecolor={dark-blue}, urlcolor={medium-blue}
}

\begin{document}

\title{High T$_C$ ferromagnetic inverse Heusler alloys: A comparative study of Fe$_2$RhSi and Fe$_2$RhGe}

\author{Y. Venkateswara}
\affiliation{Magnetic Materials Laboratory, Department of Physics, Indian Institute of Technology Bombay, Mumbai 400076, India}

\author{S. Shanmukharao Samatham}
\affiliation{Magnetic Materials Laboratory, Department of Physics, Indian Institute of Technology Bombay, Mumbai 400076, India}

\affiliation{Department of Physics, Maharaj Vijayaram Gajapathi Raj College of Engineering, Vijayaram Nagar Campus, Chintalavalasa, Vizianagaram 535005, Andhra Pradesh, India}

\author{Akhilesh Kumar Patel}
\affiliation{Magnetic Materials Laboratory, Department of Physics, Indian Institute of Technology Bombay, Mumbai 400076, India}

\author{P. D. Babu}
\affiliation{UGC-DAE Consortium for Scientific Research, Mumbai Centre, BARC Campus, Mumbai 400085, India}

\author{Manoj Raama Varma}
\affiliation{National Institute of Interdisciplinary Sciences and Technology (CSIR), Tiruvananthapuram, India}

\author{K. G. Suresh}
\email{suresh@phy.iitb.ac.in}
\affiliation{Magnetic Materials Laboratory, Department of Physics, Indian Institute of Technology Bombay, Mumbai 400076, India}

\author{Aftab Alam}
\email{aftab@phy.iitb.ac.in}
\affiliation{Materials Modeling Laboratory, Department of Physics, Indian Institute of Technology Bombay, Mumbai 400076, India}


\begin{abstract}
We report the results of experimental investigations on structural, magnetic, resistivity, caloric properties of Fe$_2$RhZ (Z=Si,Ge) along with \textit{ab-initio} band structure calculations using first principle simulations. Both these alloys are found to crystallize in inverse Heusler structure but with disorder in tetrahedral sites between Fe and Rh. Fe$_2$RhSi has saturation moment of 5.00 $\mu_B$ and while its counterpart has 5.19 $\mu_B$. Resistivity measurement reveals metallic nature in both of them. Theoretical simulations using generalized gradient approximation(GGA) predict inverse Heusler structure with ferromagnetic ordering as ground state for both the alloys. However it underestimates the experimentally observed moments. GGA+$U$ approach, with Hubbard $U$ values estimated from density functional perturbation theory helps to improve the comparison of the experimental results. Fe$_2$RhSi is found to be half metallic ferromagnet while Fe$_2$RhGe is not. Varying  $U$ values on Fe and Rh sites does not change the net moment much in Fe$_2$RhSi, unlike in Fe$_2$RhGe. Relatively small exchange splitting of orbitals in Fe$_2$RhGe compared to that of Fe$_2$RhSi is the reason for not opening the band gap in the minority spin channel in the former. High ordering temperature and moment make  Fe$_2$RhSi useful for spintronics applications. 
\end{abstract}


\date{\today}
\maketitle
\section{Introduction}

Heusler alloys (HAs)\cite{F.Heusler-1903} are well known due to their multifunctional properties such as (i) half metallic ferromagnetism,\cite{deGroot-halfmetals-prl50.2024,Katsnelson-halfmetals-RevModPhys.80.315} spin gapless semiconducting nature,\cite{Wang-SGS-PhysRevLett.100.156404,Yue-SGS-small,XTWang-SGS-JMCC} bipolar magnetic semiconductors\cite{Li-BMS-nanoscale} and spin semimetals,\cite{Venkateswara-FeRhCrGe-prb100.180404} (ii) high spin polarization (iii) supercoductivity arising in the alloys containing 27 valence electrons such as Ni$_2$ZrGa,\cite{Winterlik-ZrNi2Ga-prb78.184506} Pd$_2$RSn (R=Tb-Yb)\cite{Malik-Pd2RSn-superconductors-prb31.6971,Malik-Pd2YSn-superconductor-prb34.3144,Klimczuk-superconductors-Heusler-prb85.174505}, AuPdTM (T=Sc, Y and M=Al, Ga, In)\cite{Kautzsch-AuPdTM-MBS} etc., (iv) giant exchange bias,\cite{Nayak-exchangebias-Heusler-NM,JyothiSharma-giantexchangebias-Heusler-APL} (v) Martensitic transition which causes large magneto-caloric effect (MCE)\cite{RODIONOV-NiMnIn-GiantMCE-PP,Liu-NiMnIn-MCE-prm3.084409,Nayak-CoxNiMnSb-GiantMCE-JPDAP} and topological insulating behaviour\cite{Yang-triplepointfermion-TI-prl119.136401,Li-ATI-AB2C-prb83.235125,Pham-TI-fullHA-prb95.115124,barman-TI-ZrIrBi-prb97.075302,XTWang-X2RuPb-TI-ActaPS,Lin-halfHeusler-TI-prb91.094107,yan-halfHeusler-TI-MRSB} and Weyl semimetals.\cite{Manna-weyl-Heusler-NRM,Wang-TRBWeyl-prl117.236401} 
For the last three decades, after discovery of half-metallicity in NiMnSb by de Groot \textit{et. al.},\cite{deGroot-halfmetals-prl50.2024} HAs  gained prominence in the field of spintronics. Among the studied systems, Co-based Heusler alloys got increased attention due to their high Curie temperature ($T_C$). In addition to the $3d$ based HAs, $4d$ based alloys were also studied for spintronic applications. Some of the examples are Ru$_{2-x}$Fe$_x$CrGe,\cite{Brown-Ru2xFexCrGe-JPCM} Ru$_{2-x}$FeCrSi,\cite{SHIGETA-Ru2xFexCrSi-JPCS} Ru$_2$MnZ (Z=Si, Ge, Sn and Sb),\cite{Kanomata-Ru2MnZ-JJAP} (Ru$_{1-x}$Co$_x$)$_2$FeSi,\cite{DEKA-Ru1xCoxFeSi-PBCM} RuMn$_2$Z (Z=Si, Sn)\cite{ENDO-RuMn2Z-JAC} and CoFeRuZ (Z=Si, Ge)\cite{BAINSLA-CoRuFeX-JAC}. In these alloys, Ru couples antiferromagnetically with neighboring magnetic ions. The other important high $T_C$ series of $4d$ based Heusler alloys are Rh based. There has been extensive effort in the synthesis of the class of Rh$_2$TX (T=Ti, V, Cr, Mn, Fe, Co, NI, Cu; X=Al, Ga, Ge, Si, Sb, Pb) alloys.\cite{Suits-Rh2MnX-prb14.4131,Klaer-Rh2MnGe-JPDAP,YIN-Rh2YZ-JAC,PUGACHEVA-Rh2MnX-JMMM,KANOMATA-Rh2NiGe-JAC,PUGACHEVA-Rh2MnX-JMMM,BERRI-Rh2MnX-PBCM,EMMEL-Rh2MnGe-JMMM} However, L2$_1$ order is found only for T=Mn, Ni and Cu. Others either show tetragonal distortion or involve multiphases. Interestingly, in almost all studies in the literature on Rh-based HAs, researchers have achieved the ordered L2$_1$ structure only if the total number of valence electrons of the alloy is odd. Some of the examples include Rh$_2$MnX ( X=Ge, Sn, Pb),\cite{Suits-Rh2MnX-prb14.4131} Rh$_2$CuSn,\cite{YIN-Rh2YZ-JAC,Nikolaev-Ru2CuSn-APL,Knut-Rh2CuSn-prb88.134407} LiRh$_2$X (X=Si, Ge),\cite{BAILEY-LiRh2Si-JSSC} CoRhMnSn,\cite{Alijani-Co2xRhxMnZ-JPCM} CoRhMnGe,\cite{Rani-CoRhMnGe-prb96.184404} FeRhCrGe\cite{Venkateswara-FeRhCrGe-prb100.180404} etc. Even though Rh$_2$NiGe is reported to crystallize in L2$_1$ structure, one can notice extra impurity peaks in the reported XRD data.\cite{KANOMATA-Rh2NiGe-JAC}

Rh-based Heusler alloys, with even number of valence electrons show tetragonal distortion which can be explained using band Jahn-Teller effect.\cite{SUITS-Rh-Heusler-SSC} For these alloys, Coulomb repulsion also plays a crucial role in separating out the electronic states by broadening of the energy bands (close to the Fermi level), resulting in lattice distortion. One exception is the set of alloys containing Mn, such as CoRhMnGa,\cite{Alijani-Co2xRhxMnZ-JPCM} as Mn has the ability to adopt different oxidation states. Another reason is the nature of hybridization of Mn-atom with different neighboring orbitals as can be noticed in CoRuMnSi\cite{Venkateswara-CoRuMnSi-JMMM}, unlike the general hybridization followed in other HAs reported elsewhere.\cite{GRAF-Heusler-PSSC,Galanakis-SPrule-prb66.174429} See supplementary material\cite{Venkateswara-FeRhCrGe-prb100.180404} for more details.

In this paper, we report two new odd-valence electron Rh-based full Heusler alloys Fe$_2$RhZ (Z=Si, Ge). A detailed experimental investigation involving structural, magnetic and transport behavior is carried out. Additionally, first principle calculations are done to better understand the magnetic ordering and the electronic structure.  

\begin{figure}[h!]
\centering
\includegraphics[width=0.8\linewidth]{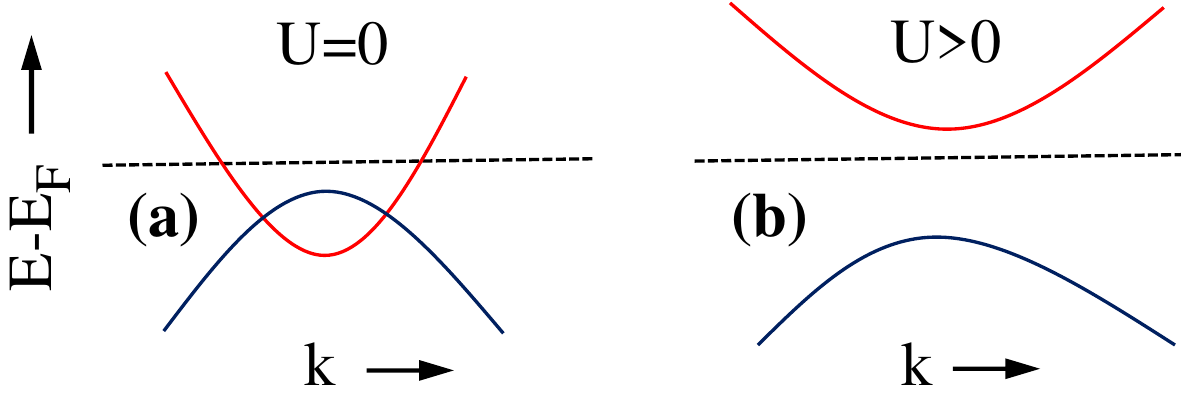}
\caption{Atom projected orbital character of bands close to Fermi level. (Left) Direct overlap of orbitals in the absence of Hubbard (U) correction, (right) gapped orbitals due to the inclusion of Hubbard correction. Partially occupied orbital (red color) corresponds to octahedral site while valence orbital (blue color) corresponds to tetrahedral sites in Heusler alloy.}
\label{fig:HubbardU}
\end{figure}

Fe$_2$RhZ (Z=Si, Ge) are 29 valence electrons Heusler systems and hence are the analogue of $3d$-based Heusler alloys such as Fe$_2$CoSi,\cite{Luo-Fe2CoSi-JPDAP} Fe$_2$CoGe,\cite{REN-Fe2CoGe-PBCM,Gasi-Fe2YZ-prb87.064411} Co$_2$FeSi,\cite{Wurmehl-Co2FeSi-APL,Wurmehl-Co2FeSi-prb-exp-theory,Sargolzaei-Co2YZ-PRB-theory,Benjamin-Co2FeZ-exptheory-STAM} Co$_2$FeGe\cite{KIM-Co2FeGe-theory-TSF,Uvarov-Fe2CoGe-thinfilm-JAP,Rai-Co2FeGe-theory-PBCM,Hyun-Co2FeGe-thinfilm-exptheory-JKPS} etc. Wurmehl \textit{et. al.}\cite{Wurmehl-Co2FeSi-prb-exp-theory,Benjamin-Co2FeZ-exptheory-STAM} and Uvarov \textit{et. al.}\cite{Uvarov-Fe2CoGe-thinfilm-JAP,Rai-Co2FeGe-theory-PBCM} studied Co$_2$FeSi and Co$_2$FeGe respectively and reported that they could only produce the experimentally observed moment and gap in the minority band by applying Hubbard $U$ on these systems. But they have neither discussed why they have to include Hubbard $U$ in their systems nor gave a clear indication to what type of Heusler systems it should be included. There is an ambiguity of inclusion of $U$ in all Heusler systems. Here we present a systematic way to observe and identify which Heusler systems need to be taken care with the inclusion of Hubbard correction. Figure \ref{fig:HubbardU}(a) represents the atom resolved orbital character of band structure without Hubbard correction. If one notices such a direct overlap of orbitals (or slightly gapped) typically around the edges of the Brillouin zone (or away from the $\Gamma$ point), they need to identify the atomic orbital character of these bands. (One can recall that the orbitals at $\Gamma$ represent molecular levels of the corresponding system). If the partially occupied conduction band (indicated by red color in Fig.\ref{fig:HubbardU}(a)) arises from the octahedral site while the valence band arises from the tetrahedral site, then one should not neglect the Hubbard correction in the Heusler system. This kind of scenario generally occurs in high valence systems in the spin down band around $X$-point in the Brillouin zone. As per the empirical rule stated in Ref.[\cite{EVS-CoFeCrGe-prb92.224413}] for the formation of Heusler alloys based on electronegativities of constituent atoms, the octahedral sites try to lose partial electrons. Hence, the partially occupied conduction band which corresponds to octahedral site lose their states by shifting their orbitals above the Fermi level. Such an observation can only be achieved by the inclusion of $U$ in the system as shown in Fig.\ref{fig:HubbardU}(b). Because $U$ correction favours integer particle numbers in the system by penalizing the partial occupancies. Therefore one should look for such direct overlap of orbitals (i) in minority band for ferro, ferri- or fully compensated ferri- magnets or (ii) in the full band structure for anti-ferromagnets or non-magnetic systems. The effect of $U$ is to open up the gap around the $k$-point in the Brillouin zone but does not always guarantee in the entire Brillouin zone as shown for Fe$_2$RhGe later in this paper. With this methodology one can predict very accurate results prior to experimental observations.

Theoretically, full Heusler alloys can accommodate a maximum of 31 valence electrons and can have a moment close to 7 $\mu_B$. This happens only when the spin up bands are completely filled with integral number of electrons. Experimentally, all such high valence electron systems reported in the literature belong to 29 and 30 valence electrons category. Fe$_2$RhZ (Z=Si, Ge) are the two new systems belonging to this category, with reasonably large $T_C$ and magnetic moment.

\section{Experimental techniques}

Both Fe$_2$RhSi and Fe$_2$RhGe alloys were prepared in polycrystalline form by arc-melting method. Room temperature X-ray diffraction (XRD) patterns were collected by PANalytical X$'$Pert Pro powder diffractometer using Cu K$_\alpha$ radiation. Rietveld refinement of XRD patterns were analyzed using FullProf\cite{FullProf-PBCM} package while VESTA\cite{Momma-VESTA-JAC} software is used for visualizing crystal structures. Magnetization measurements were carried out using  Physical Property Measurement System (PPMS) model 6000 within the vibrating sample magnetometer (VSM) option. Resistivity measurements were carried out using PPMS at different magnetic fields. Specific heat measurements were carried out from 2 to 280 K at zero field in PPMS using relaxation calorimetry.

\section{Computational details}

\begin{figure}[h]
	\centering
\subfigure[]{\includegraphics[width=0.4\linewidth]{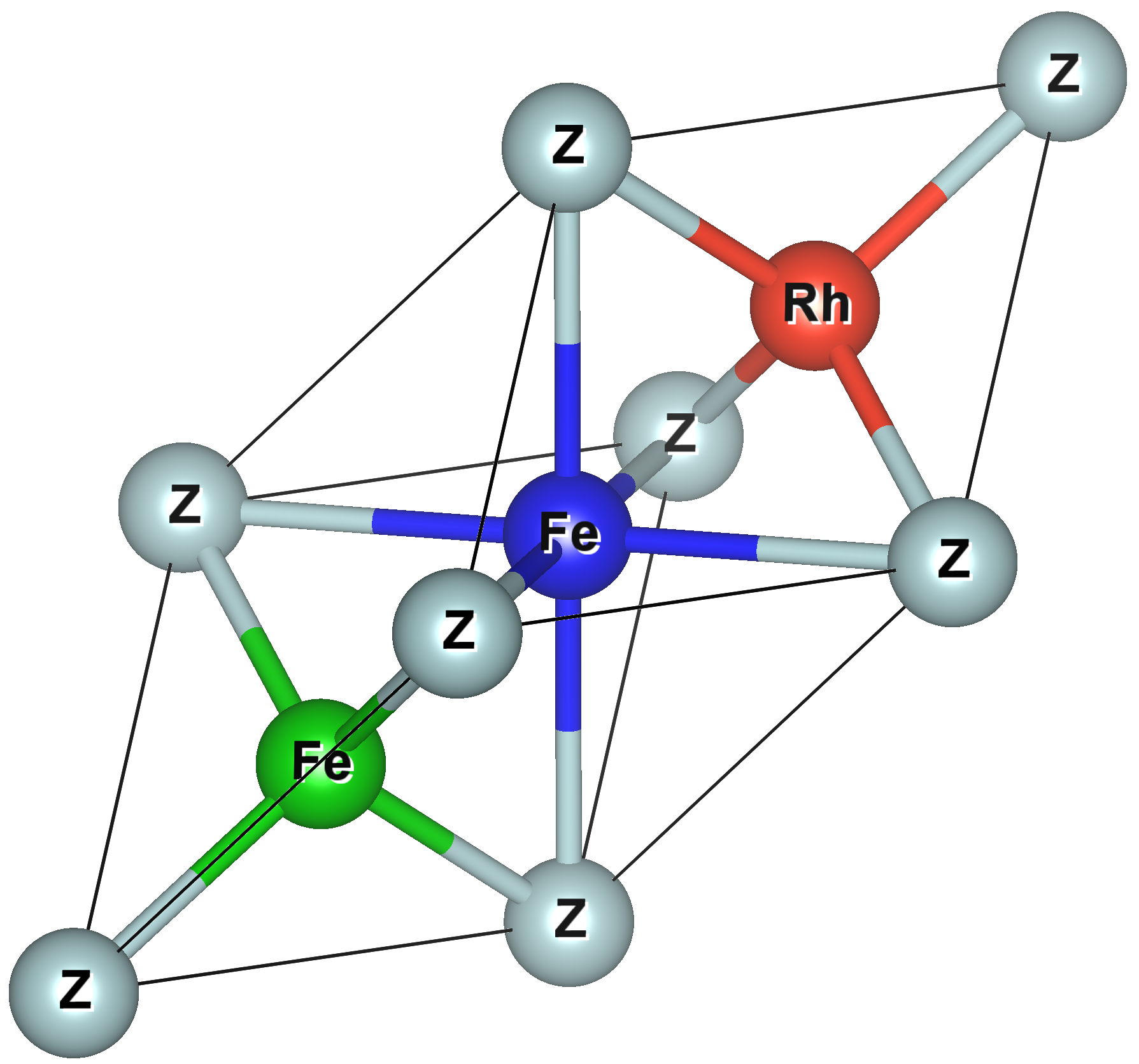}} \hspace*{0.15\linewidth}%
\subfigure[]{\includegraphics[width=0.4\linewidth]{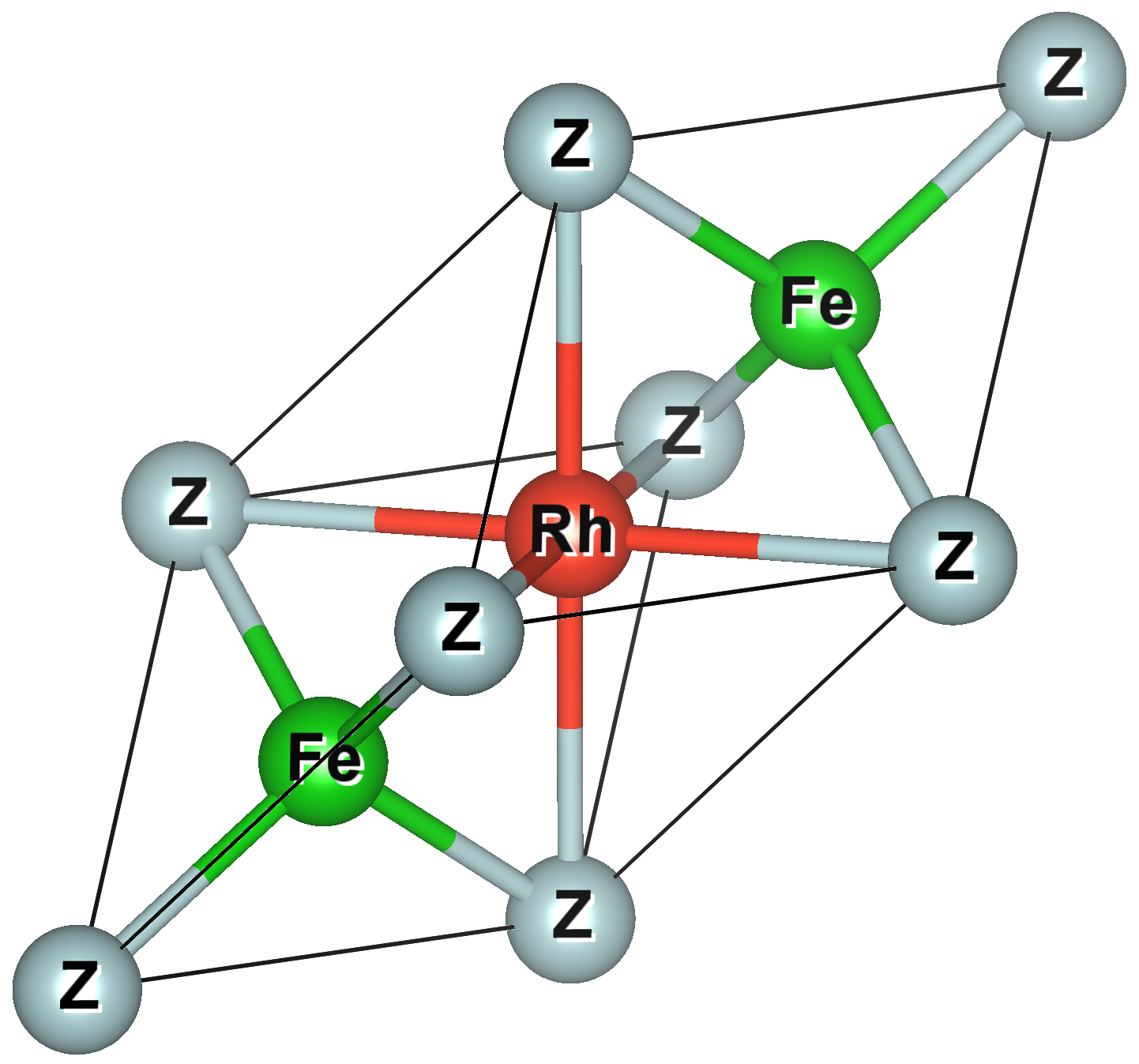}}
	\caption{Possible crystallographic configurations for Fe$_2$RhZ (Z=Si, Ge) full HAs. (a) Inverse Heusler structure (configuration I) and (b) normal Heusler structure (configuration II).}
	\label{fig:possible_conf_fe2rhz_exp}
\end{figure}

Fe$_2$RhZ (Z=Si, Ge) with the stoichiometry 2:1:1 belong to the full Heusler alloy family. There exist two non-degenerate crystal configurations/structures in this class: i) inverse Heusler structure (X type structure with prototype CuHg$_2$Ti, space group $F\bar{4}3m$) and ii) normal Heusler structure (L2$_1$ structure with prototype Cu$_2$MnAl, space group $Fm\bar{3}m$) which are shown in Figs.~\ref{fig:possible_conf_fe2rhz_exp}(a) and \ref{fig:possible_conf_fe2rhz_exp}(b) respectively. Since the valence of Fe is less than that of Rh, these alloys are expected to crystallize in inverse Heusler structure. It originates from the simple empirical rule.\cite{GRAF-Heusler-PSSC} In a full HA $X_2YZ$, keeping $Z$ at $4a$ site ( i.e. at (0,0,0) ), the lattice sites involving Fe and Rh atoms can have the following two configurations:
\begin{enumerate}[I.]
	\item Fe1 at $4b(\frac{1}{2},\frac{1}{2},\frac{1}{2})$, Fe2 at $4c(\frac{1}{4},\frac{1}{4},\frac{1}{4})$ and Rh at $4d(\frac{3}{4},\frac{3}{4},\frac{3}{4})$,
\item Rh at $4b(\frac{1}{2},\frac{1}{2},\frac{1}{2})$ and Fe's at $4c(\frac{1}{4},\frac{1}{4},\frac{1}{4})$ and $4d(\frac{3}{4},\frac{3}{4},\frac{3}{4})$.
\end{enumerate}

We used Quantum ESPRESSO\cite{QE-2017,QE-2009} to simulate the two systems with different initial structural and spin configurations on Fe and Rh ions, and find the most stable (ground) state. Exchange correlations are incorporated within the generalized gradient approximation (GGA) and the pseudopotentials parameterized by plane augmented wave method (KJPAW)\cite{KJPAW-prb-theory-pseudo} were generated using PSlibrary\cite{DALCORSO-PSlibrary-CMS}. Other computational parameters used here are the same as in our previous report.\cite{Venkateswara-CrVTiAl-prb97.054407} We used XCrySDen for making $k$-point path for band structure calculations.\cite{Xcrysden-citation}

\section{Experimental results}
\subsection{Crystal structure}

\begin{figure}[t]
	\centering
	\includegraphics[width=\linewidth]{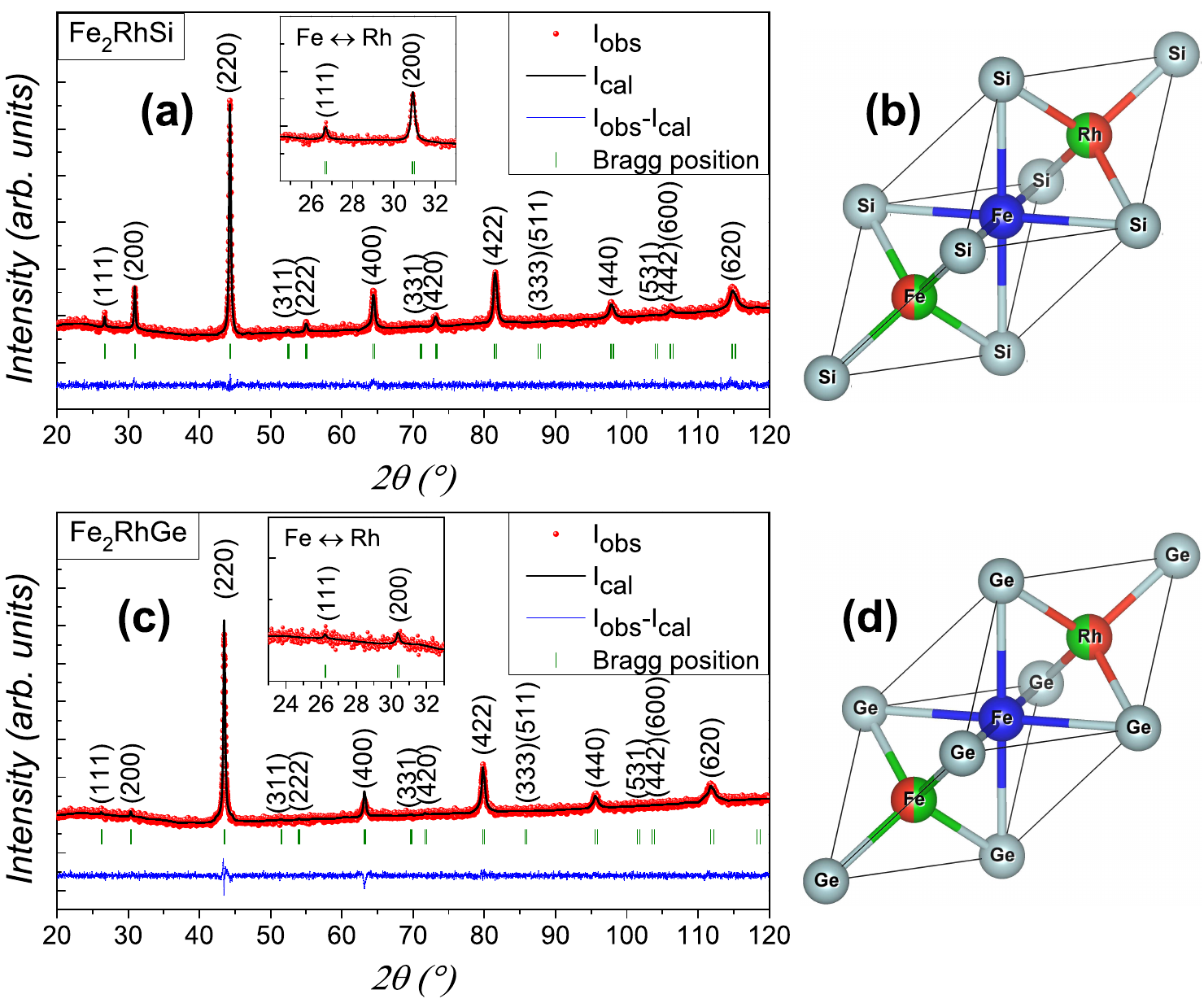}
	\caption{Rietveld refinement of room temperature XRD data for (top) Fe$_2$RhSi and (bottom) Fe$_2$RhGe respectively. Inset of (a) and (c) show the zoomed-in view of the refined data considering disorder between Fe and Rh in inverse Heusler structure (best fit). (b) and (d) Primitive cells corresponding to the best fit in (a) and (c) respectively.}
	\label{fig:XRD-Fe2RhZ}
\end{figure}

Figures \ref{fig:XRD-Fe2RhZ}(a) and \ref{fig:XRD-Fe2RhZ}(c) show the room temperature XRD pattern along with their Rietveld refinement for Fe$_2$RhSi and Fe$_2$RhGe respectively. The XRD pattern can be indexed with CuHg$_2$Ti type inverse Heusler structure with lattice parameters  5.77 \AA\ and 5.88 \AA\ for the two alloys respectively. However, the observed peak intensity is weaker than that calculated for the odd superlattice reflections such as (111), (311) \textit{etc.} indicating disorder between either tetrahedral site or octahedral site atoms. Octahedral disorder between Fe and Z(Si,Ge) does not fit well whereas 50\% disorder of tetrahedral site atoms Fe and Rh in configuration I fits very well (see the zoomed-in view of Figs. \ref{fig:XRD-Fe2RhZ}(a) and \ref{fig:XRD-Fe2RhZ}(c)). Figures \ref{fig:XRD-Fe2RhZ}(b) and \ref{fig:XRD-Fe2RhZ}(d) show the primitive cells corresponding to the best fit. Refinement does not fit well for both odd and even superlattice reflection peaks in configuration II (Cu$_2$MnAl type structure). Moreover, any amount of disorder such as (i) L2$_1$ disorder between either octahedral or tetrahedral sites, (ii) DO$_3$ disorder, (iii) B2 disorder and (iv) A2 disorder also did not fit well for configuration II. Hence, we conclude that both the alloys crystallize in inverse Heusler structure with 50\% disorder between tetrahedral site atoms Fe and Rh.

\subsection{Magnetization}

The magnetization ($M$) in full Heusler alloys can be roughly estimated using the Slater-Pauling (SP) rule,\cite{Galanakis-SPrule-prb66.174429} given below
\begin{equation}
M = (N_v - 24)\ \ \ \mu_B/f.u.,
\end{equation}
where $N_v$ is the total number of valence electrons in the alloy. Since both the alloys have 29 valence electrons, they are expected to have a saturation moment of 5 $\mu_B$/f.u., as per SP rule.

\begin{figure}[h!]
\centering
\includegraphics[width=0.9\linewidth]{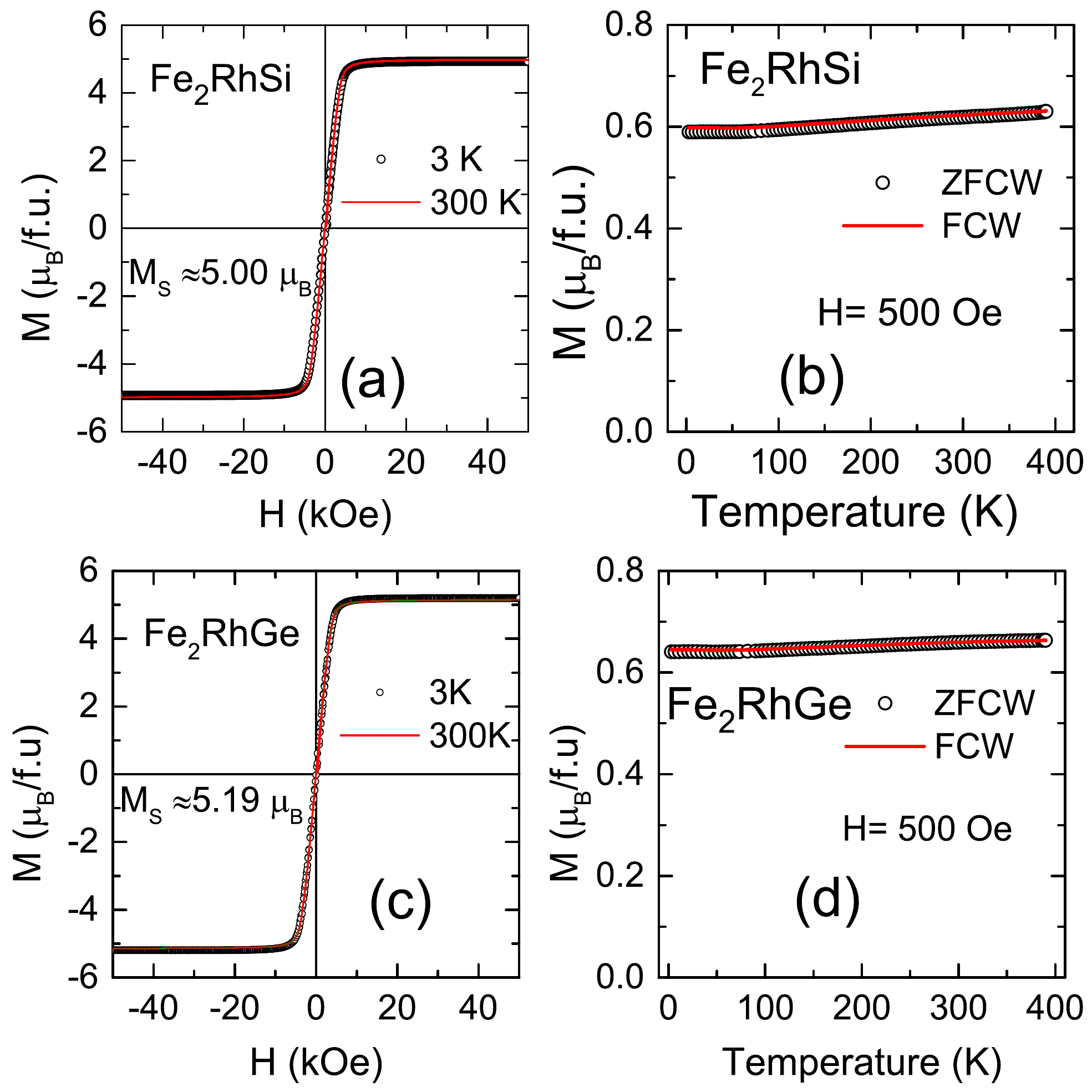}
\caption{(a) Magnetization vs. field  (H) at 3 K and 300 K for (a) Fe$_2$RhSi and (c) Fe$_2$RhGe.  Magnetization vs. temperature (T) at 500 Oe for (b) Fe$_2$RhSi and (d) Fe$_2$RhGe.}
\label{fig:mag-Fe2RhZ}
\end{figure}

Both the alloys are found to be ferromagnetic in nature with high Curie temperatures. Magnetization vs. field data at 3 K and 300 K are shown in Figs. \ref{fig:mag-Fe2RhZ}(a) and \ref{fig:mag-Fe2RhZ}(c) for Fe$_2$RhSi and Fe$_2$RhGe respectively. Fe$_2$RhSi has saturation moment of 5 $\mu_B/f.u.$, whereas Fe$_2$RhGe has 5.19 $\mu_B/f.u.$, which are consistent with the SP rule. Nearly integer moment on Fe$_2$RhSi indicates the possibility of half-metallic nature. Figure \ref{fig:mag-Fe2RhZ}(b) and \ref{fig:mag-Fe2RhZ}(d) show the temperature dependence of magnetization for the two alloys at 500 Oe. We have also performed high $T$ magnetization measurements up to 1000 K which confirm the $T_C$ for Fe$_2$RhSi and Fe$_2$RhGe to be 925 K and 910 K respectively.

\subsection{Resistivity}

\begin{figure}[b]
\centering
\includegraphics[width=\linewidth]{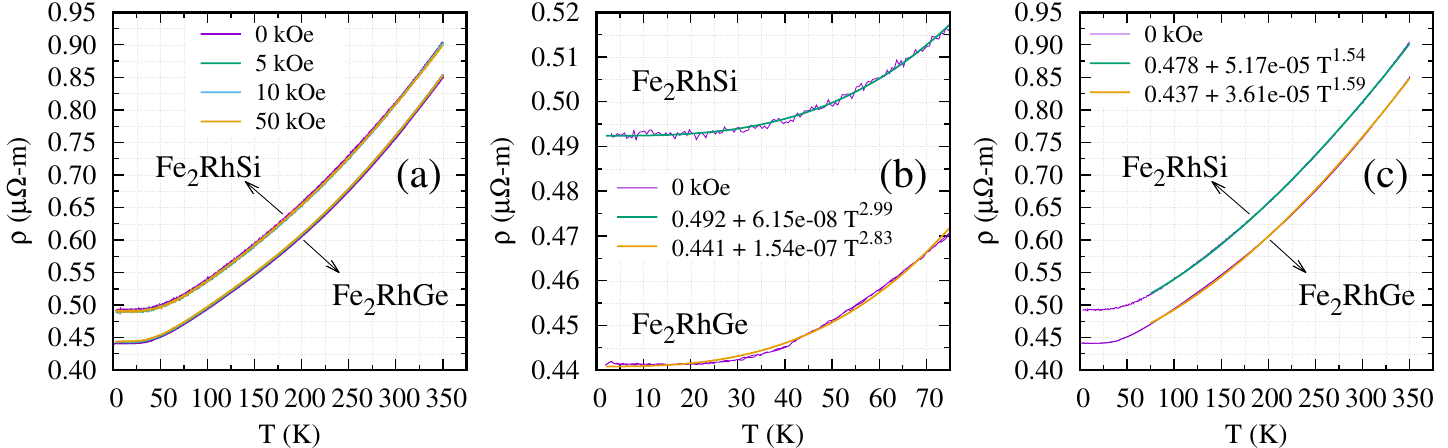}
\caption{(Color online) (a) Temperature($T$) dependence of resistivity ($\rho$) at different magnetic fields for Fe$_2$RhSi and Fe$_2$RhGe. (b) Low $T$ and (c) high $T$ exponent fit $(\rho(T)=\rho_0+AT^n)$ to the resistivity data for Fe$_2$RhSi and Fe$_2$RhGe.\\
}
\label{fig:RT-Fe2RhZ}
\end{figure}

Figure \ref{fig:RT-Fe2RhZ} shows the temperature dependance of resistivity for Fe$_2$RhSi and Fe$_2$RhGe, measured in the range 2-350 K. Measurements were carried out at different magnetic fields 0, 5, 10 and 50 kOe. In zero field, the positive temperature coefficient of $\rho(T)$ indicates metallic behavior. The application of field does not change the resistivity behavior much, as evident from Fig. \ref{fig:RT-Fe2RhZ}(a), thereby indicating the robust magnetic ordering of the alloys. The absence of positive magnetoresistance indicates the marginal effect of field on the motion of conduction electrons.  Resistivity data is fitted with the relation $\rho=\rho_0+AT^n$ where $\rho_0$ is the residual resistivity. The residual resistance ratio ($RRR$ = $R_{300 K}/R_{2 K}$) of Fe$_2$RhSi and Fe$_2$RhGe is $\approx$ 1.6. Such of a low RRR, as compared to that of conventional metals, is typical to Heusler alloys. Figures \ref{fig:RT-Fe2RhZ}(b) and \ref{fig:RT-Fe2RhZ}(c) show the fitted curve  in two temperature regions (2-75 K and 50-350 K) with almost equal residual resistivity values. In the low $T$-region, the exponents turn out be $\sim$2.99 and $\sim$2.83 for Fe$_2$RhSi and Fe$_2$RhGe respectively while they are $n\sim 1.54$ and $\sim$1.59 in the high $T$-range. Furukawa derived the expression $\rho\propto (T/D_s)^3$ for the possible anomalous single magnon scattering at low temperatures in half metals.\cite{Furukawa-JPSJ-T^3} The same behavior is also observed in half metallic systems Sm$_{0.6}$Sr$_{0.4}$MnO$_3$ and (Nd$_{0.8}$Tb$_{0.2}$)$_{0.6}$Sr$_{0.4}$MnO$_3$.\cite{Akimoto-PRL-AMS-exp} Hence the $T^3$ dependence observed in Fe$_2$RhSi at low temperatures indicates the half metallic nature whereas the exponent for Fe$_2$RhGe is slightly off from cubic dependence. The exponent in the high $T$ fitting, however is close to $5/3$, indicating the presence of high temperature spin waves.\cite{Ueda-Moriya-T^1.67-RT-JPSP,Kouacou-PtMnSb-RT-JAS} 

\subsection{Specific heat}

\begin{figure}[h!]
\centering
\includegraphics[width=\linewidth]{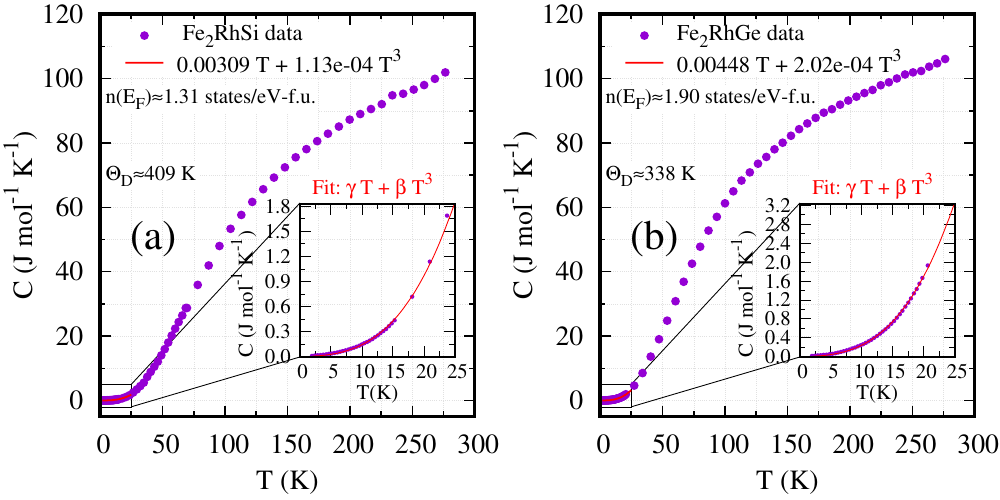}
	\caption{Specific heat(C) vs. T for (a) Fe$_2$RhSi and (b) Fe$_2$RhGe between 2-280 K in zero field. Insets show the zoomed in view of the low temperature fit, $C(T)=\gamma T+\beta T^3$ in the T-range 2-22 K.}
\label{fig:HC-Fe2RhZ}
\end{figure}

Figure \ref{fig:HC-Fe2RhZ}(a) and \ref{fig:HC-Fe2RhZ}(b) show the temperature dependence of zero field specific heat (C) for Fe$_2$RhSi and Fe$_2$RhGe respectively. The low temperature specific heat is fitted to the expression $C(T)=\gamma T+\beta T^3$ from 2-22 K, as shown in insets of Figs. \ref{fig:HC-Fe2RhZ}(a) and \ref{fig:HC-Fe2RhZ}(b) respectively. Here $\gamma T$ and $\beta T^3$ are electronic and low temperature phonon contributions. The density of states (DOS) at the Fermi level $n(E_F)$ responsible for electronic contribution is estimated using the relation $n(E_F)=3\gamma/(\pi k_B)^2$,\cite{Tseng-PRB-Ru2TaAl-semimetal} where $\gamma$ is Sommerfield constant obtained from the fit and $k_B$ is Boltzmann constant. The Debye temperature $\Theta_D$ is estimated using the relation $\Theta_D=\sqrt[3]{1944p/\beta}$, where $\beta$ is the coefficient of $T^3$ dependence at low temperatures and $p$ is the number of atoms in a formula unit. The values of $\gamma$ and $\beta$ in units of $mJ\ mol^{-1} K^{-2}$ and $J\ mol^{-1} K^{-4}$ are estimated to be 3.09$\pm$0.33 and (1.13$\pm$0.01)$\times10^{-4}$ for Fe$_2$RhSi and 4.48$\pm$0.49, (2.02$\pm$0.02)$\times 10^{-4}$ for Fe$_2$RhGe. Thus estimated $n(E_F)$ are 1.31$\pm$0.13 states/eV-f.u. and 1.90$\pm$0.21 states/eV-f.u. for Fe$_2$RhSi and Fe$_2$RhGe respectively. The estimated $\Theta_D$ are 409 K and 338 K for Fe$_2$RhSi and Fe$_2$RhGe respectively. The estimated $n(E_F)$ values are in good agreement with the simulated results using GGA+$U$ approach. It predicts nearly 1.0 states/eV-f.u. and 1.9 states/eV-f.u. for Fe$_2$RhSi and Fe$_2$RhGe respectively. The values of $C$ slightly larger than 100 $J\ mol^{-1}K^{-1}$ (Dulong-petit law) for $T>250$ K indicate the presence of small but non-negligible magnon contribution to specific heat.

\section{Theoretical results and discussion}

\begin{table}[t]
	\centering
	\caption{Relaxed lattice parameter ($a_0$), atom projected moments and total cell moment, relative energy ($\Delta E$) of different configurations of Fe$_2$RhZ (Z=Si, Ge) within GGA functional.}
	\begin{tabular}{l c c c c c c c }
		\hline \hline 	
		\multirow{3}{*}{Config. }& \multirow{3}{*}{Alloy} & \multirow{3}{*}{$a_0$ (\AA)} & \multicolumn{4}{c}{Moment ($\mu_B$)}  &  {$\Delta E$} \\
		& & & & & & \\
		&		&			 	&	4d		&	4b	&	4c		&	Total & ($meV/atom$)	\\ \hline
		&		&			&	$\mathrm{Fe_1}$	&$\mathrm{Fe_2}$	&	$\mathrm{Rh}$ & 	& \\
		
		&$\mathrm{Fe_2RhSi}$&	5.79&		1.70			&	2.80			&		0.40		& 4.90	& 	0.0 \\
		
		I				&			&		&					&				&				&	& 	\\
		
		&$\mathrm{Fe_2RhGe}$&	5.90	&		1.84			&	2.84			&		0.35		& 5.03	& 0.0	\\

		&			&		&	     			&		    		&		    		&	& 	\\ \hline
		&			&		& $\mathrm{Fe_1}$	& $\mathrm{Rh}$	&	$\mathrm{Fe_2}$	& 	&\\
		
		&$\mathrm{Fe_2RhSi}$& 5.80	&		1.89			&	0.60			&		1.89	&	4.38	& 282\\
		
		II				&			&		&					&				&			&		& 	\\
		
		&$\mathrm{Fe_2RhGe}$& 5.91	&		2.10			&	0.62			&		2.10		& 4.92	& 235	\\ 
		
		\hline \hline
		
	\end{tabular}
	
	\label{tab:LDA_GGA_Fe2RhZ_theory}
\end{table}

Both Fe$_2$RhSi and Fe$_2$RhGe alloys were fully relaxed in the two configurations I and II using GGA functional. The magnetic state and their total energies at the relaxed lattice parameters ($a_0$) are listed in Table \ref{tab:LDA_GGA_Fe2RhZ_theory}. Configuration I is found to be energetically, more stable indicating that the alloys prefer inverse Heusler structure. The calculated net magnetic moments for Fe$_2$RhSi and Fe$_2$RhGe are 4.90 $\mu_B$ and 5.03 $\mu_B$ respectively which are in fair agreement with SP rule and the experimental values.

\begin{figure}[b]
	\centering
	\includegraphics[width=1.0\linewidth]{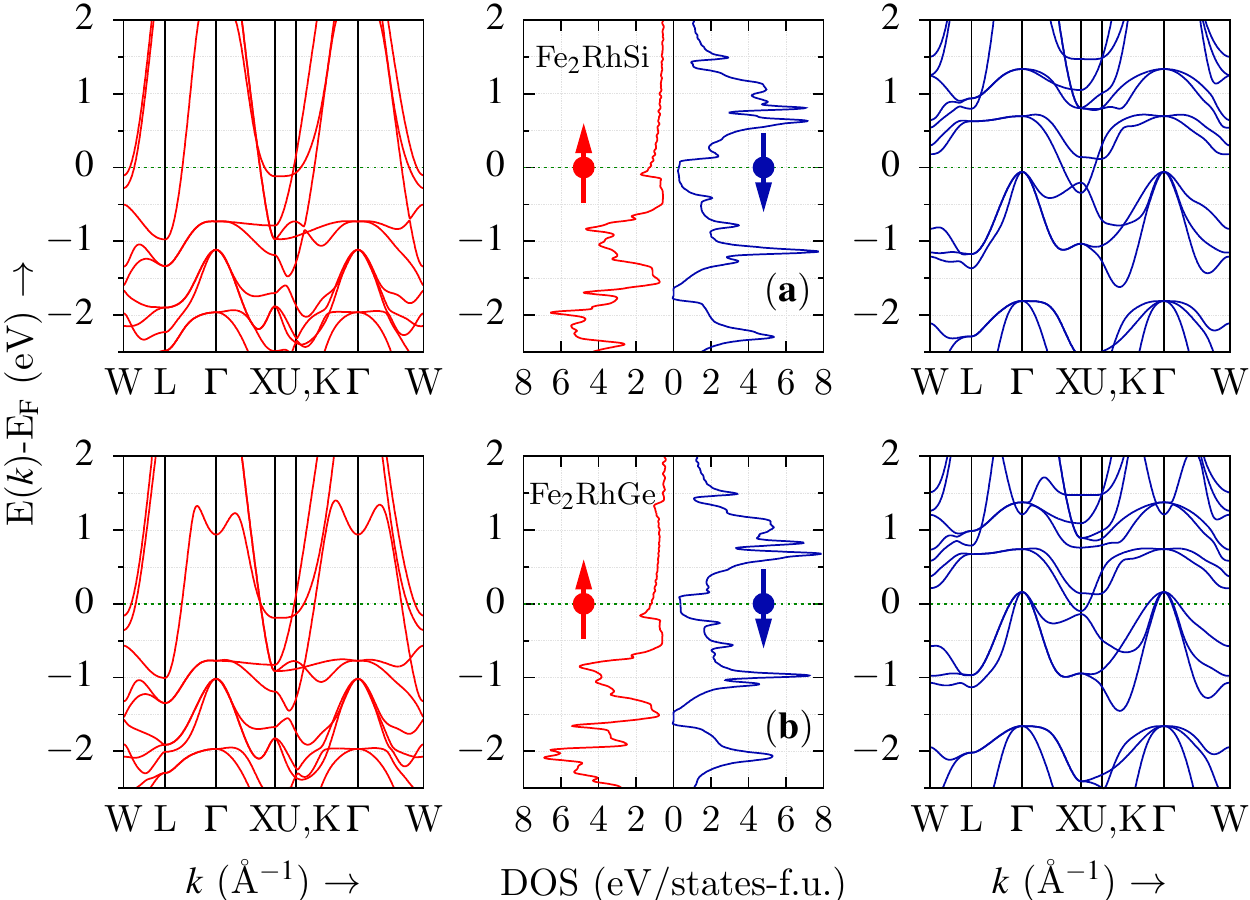}
	\caption{Spin resolved band structure and density of states for configuration I of (top) Fe$_2$RhSi and (bottom) Fe$_2$RhGe at their equilibrium lattice parameter($a_0$) within GGA functional.}
	\label{fig:DB-Fe2RhZ-GGA}
\end{figure}

\begin{figure}[t]
	\centering
	\includegraphics[width=\linewidth]{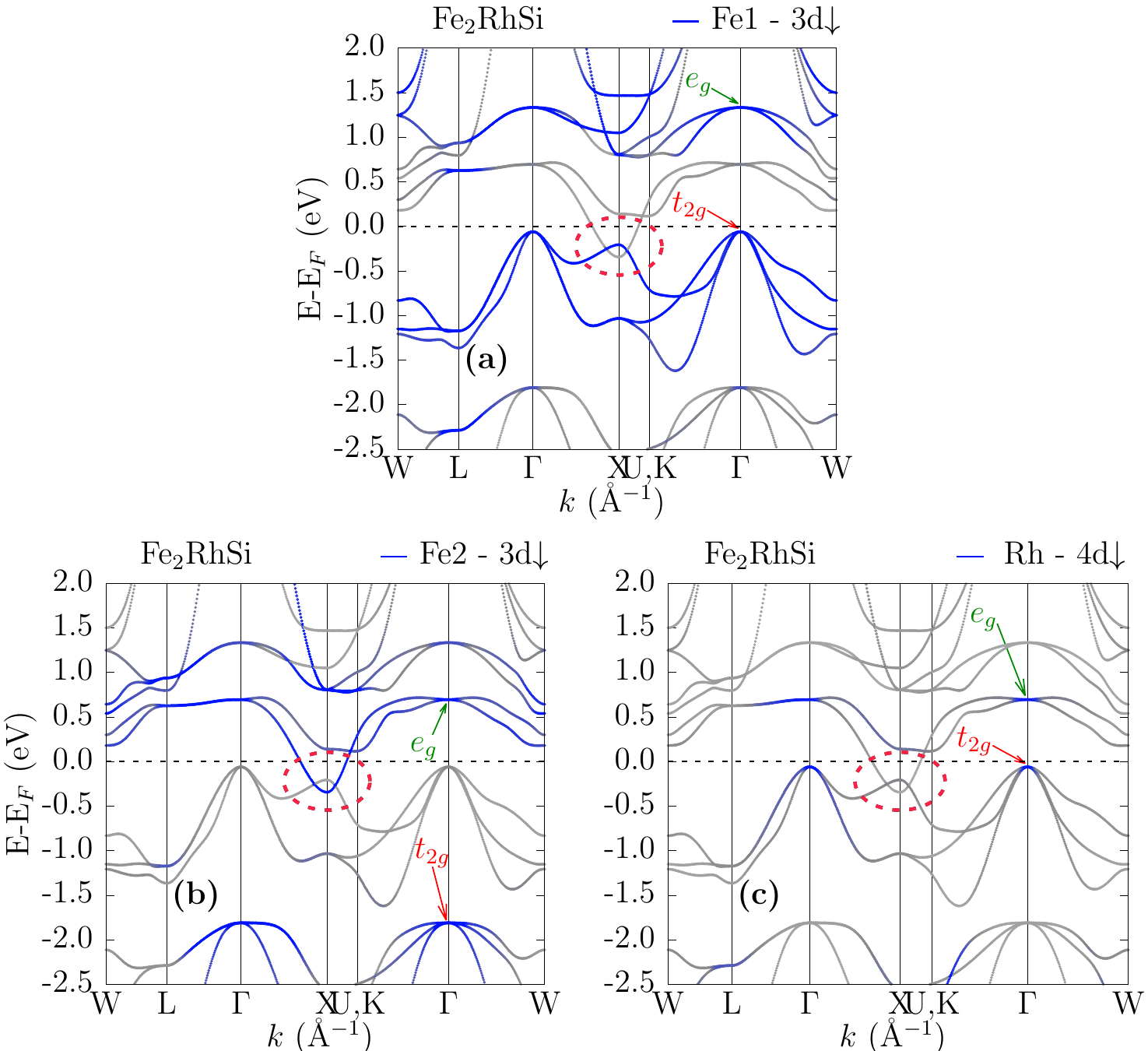}
	\caption{$d$ orbital projected band structure of spin down channel for configuration I of Fe$_2$RhSi with GGA functional. (a) $3d$ orbital character of tetrahedral site atom Fe, (b) $3d$ orbital character of octahedral site atom Fe and (c) $4d$ orbital character of another tetrahedral site atom Rh.}
	\label{fig:Fe1_Fe2_down_orbproj_Fe2RhSi1_eq}
\end{figure}

Figure \ref{fig:DB-Fe2RhZ-GGA} shows the spin resolved band structure and density of states for configuration I of Fe$_2$RhZ (Z=Si and Ge) using GGA functional. The band structure clearly indicates that neither of the two systems is half metallic due to presence of small finite DOS at $E_F$ in the spin down band. The small DOS arises due to the overlap of pair of valence and conduction bands around $X$ point in Brillouin zone. The character of these bands is obtained by projecting atomic orbitals on different atomic sites, as shown in Fig.~\ref{fig:Fe1_Fe2_down_orbproj_Fe2RhSi1_eq} for Fe$_2$RhSi. One can notice from Figs. \ref{fig:Fe1_Fe2_down_orbproj_Fe2RhSi1_eq}(a)-(c) that the conduction orbital is mainly contributed by $e_g$ character of octahedral site Fe (denoted as Fe2) while valence orbital by $t_{2g}$ character of tetrahedral site Fe (denoted as Fe1) around the $X$-point, as highlighted by dotted encircle. This depicts the Hubbard picture as these orbitals are arising from different sites, but by the same atom (i.e. Fe) and are overlapping with same moment and energy due to direct overlap. This leaves the possibility of exchange of indistinguishable particles between two distinct sites of Fe. A similar scenario can be noticed in the band structure of Fe$_2$RhGe using GGA functional (please see supplement\cite{Fe2RhZ-supplement}). Hence, we decided to carry out the simulations using GGA+$U$ functional. This is carried out in two ways, (i) self consistent evaluation of $U$ values on Hubbard atoms, i.e. Fe1, Fe2 and Rh, unlike trial and error method, (ii) step wise increment of $U$ on Hubbard atoms. 

Self consistent estimation of $U$ values on different atomic sites (Fe1, Fe2 and Rh) in both the configurations of Fe$_2$RhSi and Fe$_2$RhGe is carried out using linear response method as described by Coccioni \textit{et. al.}\cite{Cococcioni-linearresponseU-prb} Please see supplementary material\cite{Fe2RhZ-supplement} for details on deriving $U$ values at different Hubbard sites. $U$ values are estimated with increasing supercell size and the converged values are taken for the final GGA+$U$ calculations. 

\begin{table} [t]
	\caption{Self consistently converged Hubbard energies(U) on Fe1, Fe2 and Rh atoms for the two configurations (I and II) of Fe$_2$RhZ (Z=Si,Ge).}
	\begin{tabular}{l c c c c }
		
		\hline \hline 
		\multirow{3}{*}{Config.}	    &	\multirow{3}{*}{Alloy}				&	\multicolumn{3}{c}{Hubbard U (eV)}	\\ 
		&					&	\ \ \ \ \ 4d \ \ \ \ \		& \ \ \ \ \	4b \ \ \ \ \	& \ \ \ \ \	4c	\ \ \ \ \			\\ \hline
		&		&	$\mathrm{Fe1}$	&	$\mathrm{Fe2}$	&	$\mathrm{Rh}$	 \\
		
		&$\mathrm{Fe_2RhSi}$	&	5.2		&	4.1		&	6.8				\\ 
		
		I				&		&		&		&		 \\
		
		&$\mathrm{Fe_2RhGe}$	&	4.6		&	3.9		&	6.6				\\ 
		
		&		&		&		&		 \\ \hline
		&		&	$\mathrm{Fe_1}$	&	$\mathrm{Rh}$	&	$\mathrm{Fe_2}$	 \\
		
		&$\mathrm{Fe_2RhSi}$	&	3.8		&	5.3		&	3.8				\\ 
		
		II				&		&		&		&		 \\
		
		&$\mathrm{Fe_2RhGe}$	&	3.5	&	5.2	&	3.5			\\ 
		
		\hline \hline
		
	\end{tabular}
	\label{tab:U-values-Fe2RhZ-GGA}
\end{table}

Table \ref{tab:U-values-Fe2RhZ-GGA} shows the converged Hubbard $U$ energies on different atoms for both the configurations of the two systems. One can notice the different $U$ values on Fe1 and Fe2 due to their different chemical environments with $U_{Fe1}>U_{Fe2}$ for the configuration I. In the second configuration, both Fe1 and Fe2 occupy the tetrahedral sites and even if they are treated differently in our simulations, their U values came out to be same as their chemical environments are identical. The U value on Rh decreases from tetrahedral site (as in configuration I) to octahedral site (in configuration II). Slightly lower $U$-values observed in Fe$_2$RhGe as compared to Fe$_2$RhSi can be attributed to the reduced hybridization strength due to the ligand atom (i.e., Si or Ge) (stronger hybridization leads to larger band splittings as seen in Fe$_2$RhSi).

Looking at the $U$ values on Rh ($>$6.6 eV), Fe1 \& Fe2 ($>$3.9 eV) for the configuration I of Fe$_2$RhZ (Z=Si, Ge), one may argue that the estimated values are relatively large. However, these $U$-values are comparable to those of a similar, large moment system Co$_2$FeSi.\cite{Wurmehl-Co2FeSi-prb-exp-theory} This system has 30 valence electrons and hence carries a net moment of 6 $\mu_B$ according to the SP rule. Wurmehl \textit{et al.} reported it to be a ferromagnet with $T_C\approx 1100$ K and an experimental moment of 6 $\mu_B$.\cite{Wurmehl-Co2FeSi-prb-exp-theory} It was reported that the experimentally observed moment can only be reproduced by the application of U in excess of 7.5 eV.\cite{Katsnelson-halfmetals-RevModPhys.80.315} They also reported that the application of U$_{eff}$=U-J (where J is the exchange parameter) ranging between 2.5 eV to 5.0 eV on Co atom and (simultaneously) 2.4 eV to 4.8 eV on Fe atom result in a moment of 6 $\mu_B$ and a gap in the minority state.\cite{Wurmehl-Co2FeSi-prb-exp-theory} $3d$ transition elements Fe and Ni are reported to have $U$-values greater than 4.5 eV in FeO and NiO.\cite{Cococcioni-linearresponseU-prb} Therefore, the listed $U$ values in Table \ref{tab:U-values-Fe2RhZ-GGA} are within the expected range, for the considered elements.

\begin{figure}[t]
	\centering
	\includegraphics[width=\linewidth]{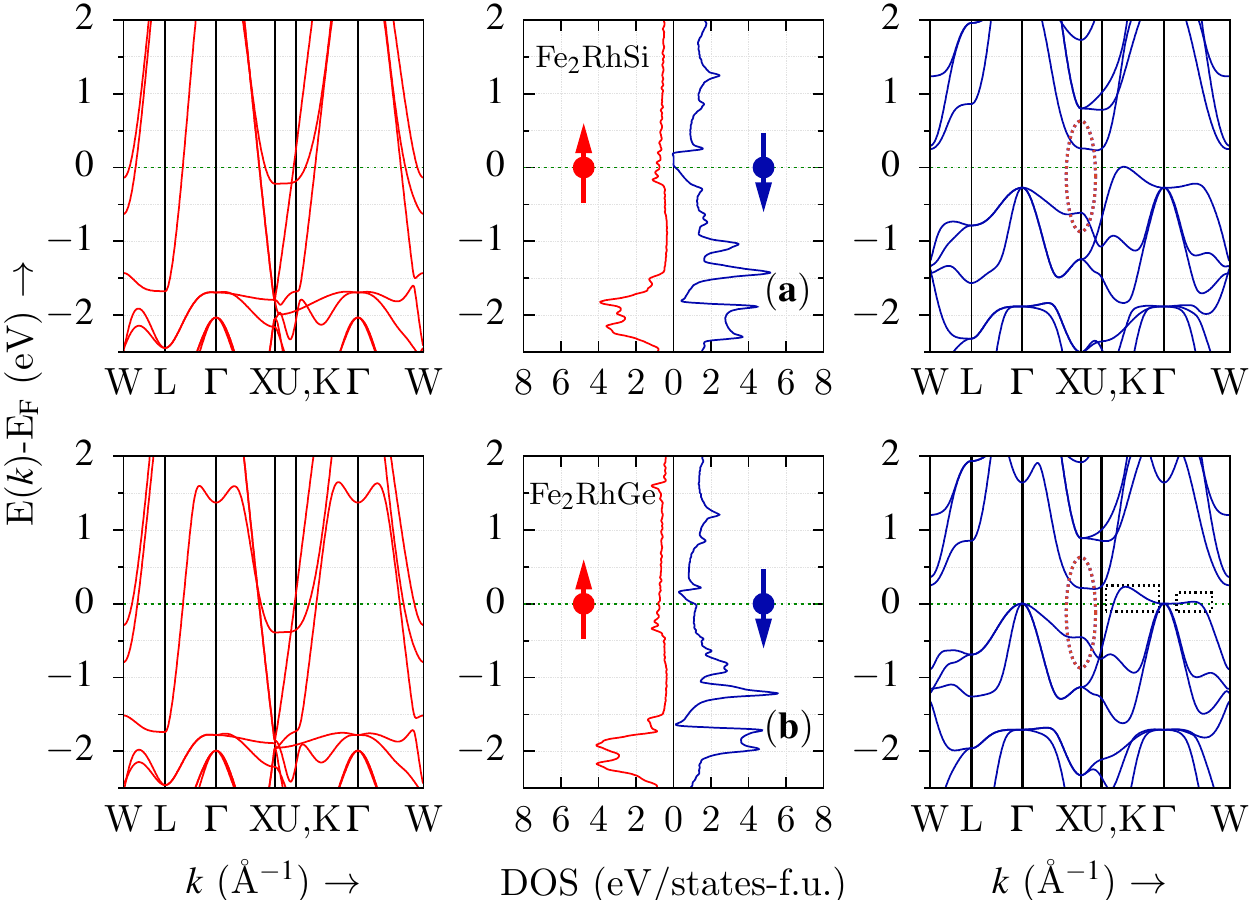}
	\caption{Spin resolved band structure and density of states for (top) Fe$_2$RhSi and (bottom) Fe$_2$RhGe alloys using GGA+$U$ approach.\cite{Cococcioni-linearresponseU-prb} The $U$-values used in this calculation are listed in Table \ref{tab:U-values-Fe2RhZ-GGA}. Dotted encircle shows the location of opening of gap. Dotted rectangle shows the location of hole pockets.}
	\label{fig:DB-Fe2RhZ-GGAU}
\end{figure}

Using the estimated $U$ values from the linear response method (as tabulated in Table \ref{tab:U-values-Fe2RhZ-GGA}) in the GGA+$U$ scheme,\cite{Cococcioni-linearresponseU-prb} electronic/magnetic properties of configurations I and II of Fe$_2$RhZ (Z=Si, Ge) were investigated. Configuration I turns out to be energetically more stable in both the cases. Figure \ref{fig:DB-Fe2RhZ-GGAU} shows the spin resolved band structure and density of states for  configuration I of Fe$_2$RhZ (Z=Ge,Si). Interestingly, Fe$_2$RhSi becomes half metal whereas Fe$_2$RhGe remains metallic. The direct overlap of $e_g$ conduction orbital and $t_{2g}$ valence orbital around $X$-point within GGA only approach (without $U$) are now separated out (see the dotted encircle region in Fig.~\ref{fig:DB-Fe2RhZ-GGAU}) due to Hubbard energies. This also changes the partially occupied conduction $e_g$ orbitals to fully unoccupied ones by shifting them above the Fermi level in Fe$_2$RhZ (Z=Si,Ge). Fe$_2$RhGe does not become half metal even after applying $U$ values because of lower band splitting energies and the presence of hole pockets in its spin down band structure. This is true even if one varies $U$ values (on Fe1, Fe2 and Rh) in  any range. On the other hand, Fe$_2$RhSi turns to be a half metal for any $U$ values above 1.0 eV on Fe1, Fe2 and Rh. One can notice the hole pockets along the paths $\Gamma-K$ and $\Gamma-W$ in the spin down band structure of Fe$_2$RhGe (see Fig.~\ref{fig:DB-Fe2RhZ-GGAU}). These hole pockets are responsible not only for causing large   total density of states at Fermi level, as evident from experimental specific heat analysis, but also for the high temperature single magnon scattering observed in resistivity analysis. The origin of high temperature single magnon contribution to resistivity, observed for Fe$_2$RhSi might be  due to the presence of hole like bands touching the Fermi level in spin down band structure (see Fig.~\ref{fig:DB-Fe2RhZ-GGAU}). This approach predicts an almost constant moment of 5.00 $\mu_B$ for Fe$_2$RhSi, unlike for Fe$_2$RhGe where the moment varies from 5.03 $\mu_B$ to 5.25 $\mu_B$ as the $U$ values increase.  

\section{Conclusion}

Fe$_2$RhZ (Z=Si, Ge) were synthesized experimentally and found to crystallize in inverse Heusler structure. Both the alloys are found to be ferromagnets with saturation magnetization 5.00 $\mu_B$ and 5.19 $\mu_B$ respectively and Curie temperature above 900 K. Resistivity measurement reveals metallic nature for both the systems. Integer moment and presence of anomalous single magnon contribution to low temperature resistivity in Fe$_2$RhSi indicate the possibility of being a half metal. Heat capacity analysis predicts larger density of states at Fermi level for Fe$_2$RhGe as compared to Fe$_2$RhSi and are in good agreement with the simulated values obtained using GGA+$U$ functional. \textit{Ab-initio} calculations using GGA approach predicts inverse Heusler structure with ferromagnetic ordering to be energetically more favorable. The band structure obatined with GGA calculations suggest metallic behavior for both the alloys. In contrast, GGA+$U$ approach (with self-consistently calculated $U$-values) opens up a gap at/around $X$-point for the directly overlapped conduction $e_g$ orbital and a valence $t_{2g}$ orbital in spin down band structure for both the systems. Interestingly it predicts a net moment of 5.00 $\mu_B$ with a half metallic nature for Fe$_2$RhSi.
Fe$_2$RhGe, however, does not become half metal because of relatively low band splitting energies (as compared to Fe$_2$RhSi) arising out of  a weak hybridization and presence of hole pockets along $\Gamma-K$ and $\Gamma-W$ path in the spin down band structure. Irrespective of $U$ values (above 1.0 eV on Fe1, Fe2 and Rh) and the nature of functionals (either LDA+$U$ or GGA+$U$), Fe$_2$RhSi is found to be a half metal while Fe$_2$RhGe remains metallic. Simulated results based on GGA+$U$ approach gives a very good overall agreement with experiment. Therefore, we conclude that Fe$_2$RhSi is a potential material for spintronics application due to its high transition temperature, half metallic nature and higher crystal stability.


\section{Acknowledgments}

YV and SSS acknowledge the financial support provided by IIT Bombay. YV acknowledges Dr. Durgesh Singh for his assistance in experimental measurements. AA acknowledges DST-SERB (Grant No. CRG/2019/002050) for funding to support this research.


\bibliographystyle{apsrev4-1}
\bibliography{bib}

\end{document}